\documentclass[12pt]{article}
\usepackage{times}

\begin{document}

\title{\textbf{\ On the solutions of the} \textbf{Schr\"{o}dinger equation with
some molecular potentials: wave function ansatz }}
\author{Sameer M. Ikhdair\thanks{%
sikhdair@neu.edu.tr} \ and \ Ramazan Sever\thanks{%
sever@metu.edu.tr} \\
{\small \textsl{$^{\ast }$Department of Physics, \ Near East University,
Nicosia, North Cyprus, Mersin-10, Turkey }} \\
{\small \textsl{$^\dagger$Middle East Technical University,Department of
Physics,06531 Ankara, Turkiye }}}
\date{\today}
\maketitle

\begin{abstract}
Making an ansatz to the wave function, the exact solutions of the $D$%
-dimensional radial Schr\"{o}dinger equation with some molecular potentials
like pseudoharmonic and modified Kratzer potentials are obtained. The
restriction on the parameters of the given potential, $\delta $ and $\eta $
are also given, where $\eta $ depends on a linear combination of the angular
momentum quantum number $\ell $ and the spatial dimensions $D$ and $\delta $
is a parameter in the ansatz to the wave function. On inserting $D=3$, we
find that the bound state eigensolutions recover their standard analytical
forms in literature.


PACS\ number: 03.65.-w; 03.65.Fd; 03.65.Ge.
\end{abstract}


\section{Introduction}

\noindent The solution of the fundamental dynamical equations is an
interesting phenomenon in many fields of physics and chemistry. To obtain
the exact $\ell $-state solutions of the Schr\"{o}dinger equation (SE) are
possible only for a few potentials and hence approximation methods are used
to obtain their solutions. According to the Schr\"{o}dinger formulation of
quantum mechanics, a total wave function provides implicitly all relevant
information about the behaviour of a physical system. Hence if it is exactly
solvable for a given potential, the wave function can describe such a system
completely. Until now, many efforts have been made to solve the stationary
SE with anharmonic potentials in three dimensions and two dimensions [1-5]
with many applications to molecular and chemical physics. The study of the
SE with these potentials provides us with insight into the physical problem
under consideration. However, the study of SE with some of these potentials
in the arbitrary dimensions $D$ is presented in (cf. Ref.[6] and the
references therein). Furthermore, the study of the bound state processes is
also fundamental to understanding of molecular spectrum of a diatomic
molecule in quantum mechanics [7]. The purpose of this paper is to carry out
the analytical solutions of the $D$-dimensional radial SE with some diatomic
molecular potentials by making a suitable ansatz to the wave function. These
molecular potentials can be taken as pseudoharmonic potential [8,9] $%
V(r)=D_{e}\left( \frac{r}{r_{e}}-\frac{r_{e}}{r}\right) ^{2},$ and a
modified Morse or Kratzer-Fues potential [10-15] which is of a Mie-type form
(cf. Ref. [14] and the references therein) $V(r)=D_{e}\left( \frac{r-r_{e}}{r%
}\right) ^{2}$ where $D_{e}$ is the dissociation energy between two atoms in
a solid and $r_{e}$ is the equilibrium intermolecular seperation. Morse
potential [10] is commonly used for anharmonic oscillator. However, its
wavefunction is not vanishing at the origin. On the other hand, the Mie-type
and also the pseudoharmonic potentials [8,9] do vanish. The Mie-type
potential [14] has the general features of the true interaction energy [16],
interatomic and dynamical properties in solid-state physics [17]. The
pseudoharmonic potential may be used for the energy spectrum of linear and
non-linear systems [18]. The pseudoharmonic [8,9] and Mie-type [13-15]
potentials are two exactly solvable potentials other than the Coulombic and
anharmonic oscillator [9]. This paper is organized as follows. In Section
\ref{TDD}, we solve analytically the $D$-dimensional radial Schr\"{o}dinger
equation with the molecular pseudoharmonic and modified Kratzer-Fues
potentials. We also obtain the bound state eigensolutions of these molecular
diatomic potentials by making a suitable ansatz to every wave function. The
results are used to find energy spectra of diatomic $N_{2},$ $CO,$ $NO$ and $%
CH$ molecules [15,19-20]. The concluding remarks will be given in Section
\ref{CR}.

\section{The $D$-Dimensional Radial Schr\"{o}dinger Equation with Some
Molecular Potentials}

\label{TDD}In the $D$-dimensional Hilbert space, the radial wave function $%
\psi (r)$ of the Schr\"{o}dinger equation can be written as [21]
\begin{equation}
\left[ \frac{d^{2}}{dr^{2}}+\frac{\left( D-1\right) }{r}\frac{d}{dr}-\frac{%
\ell (\ell +D-2)}{r^{2}}+\frac{2\mu }{\hbar ^{2}}\left( E-V(r)\right) \right]
\psi (r)=0,
\end{equation}
where $\ell $ denotes the angular momentum quantum number and $\mu $ is the
reduced mass. Furthermore, to remove the first derivative from the above
equation, we define a new radial wave function $R(r)$ by means of equation
\begin{equation}
\psi (r)=r^{-(D-1)/2}R(r),
\end{equation}
which will give the radial wave function $R(r)$ satisfying the wave equation
\begin{equation}
\left\{ \frac{d^{2}}{dr^{2}}-\frac{\left[ 4\ell \left( \ell +D-2\right)
+\left( D-1\right) \left( D-3\right) \right] }{4r^{2}}+\frac{2\mu }{\hbar
^{2}}\left( E-V(r)\right) \right\} R(r)=0.
\end{equation}
Further, Equation (3) can be written into a simple analogy of the
two-dimensional radial Schr\"{o}dinger equation as [2,6]
\begin{equation}
\left\{ \frac{d^{2}}{dr^{2}}+\frac{2\mu }{\hbar ^{2}}\left( E-V(r)\right) -%
\frac{\left( \eta ^{2}-1/4\right) }{r^{2}}\right\} R(r)=0,
\end{equation}
with the parameter
\begin{equation}
\eta =\ell +\frac{1}{2}\left( D-2\right) ,
\end{equation}
which is a linear combination of the spatial dimensions $D$ and the angular
momentum quantum number $\ell .$

\subsection{The pseudoharmonic potential}

This potential has been studied in three dimensions [8] using the polynomial
solution method. It has the following form [8]
\begin{equation}
V(r)=D_{e}\left( \frac{r}{r_{e}}-\frac{r_{e}}{r}\right) ^{2},
\end{equation}
which can be simply rewritten in the form of isotropic harmonic oscillator
plus inverse quadratic potential [9] as
\begin{equation}
V(r)=ar^{2}+\frac{b}{r^{2}}+c,
\end{equation}
where $a=D_{e}r_{e}^{-2},$ $b=D_{e}r_{e}^{2}$ and $c=-2D_{e}.$ We want to
solve the radial SE, \ Eq. (4), with the potential given in Eq. (7) by
taking an ansatz for the radial portion of wave function
\begin{equation}
R(r)=\exp \left[ p\left( \alpha ,r\right) \right] \sum_{n=0}a_{n}r^{2n+%
\delta +3/2},
\end{equation}
where
\begin{equation}
p\left( \alpha ,r\right) =\frac{1}{2}\alpha r^{2}.
\end{equation}
Substituting Eq. (8) into Eq. (4) and equating the coefficient of $%
r^{2n+\delta +3/2}$ to zero, one can obtain
\begin{equation}
A_{n}a_{n}+B_{n+1}a_{n+1}+C_{n+2}a_{n+2}=0,
\end{equation}
where
\begin{equation}
A_{n}=\frac{2\mu }{\hbar ^{2}}\left( E-c\right) +2\alpha \left( 2n+\delta
+2\right) ,
\end{equation}
\begin{equation}
B_{n}=-\frac{2\mu b}{\hbar ^{2}}-\left( \eta ^{2}-\frac{1}{4}\right) +\left(
2n+\delta +\frac{3}{2}\right) \left( 2n+\delta +\frac{1}{2}\right) ,
\end{equation}
\begin{equation}
C_{n}=0,
\end{equation}
and the value of the parameter for $p\left( \alpha ,r\right) $ can be
evaluated as
\begin{equation}
\alpha =\pm \sqrt{\frac{2\mu a}{\hbar ^{2}}}.
\end{equation}
To obtain a well-behaved solution at the origin and infinity, we must set $%
\alpha =-\sqrt{\frac{2\mu a}{\hbar ^{2}}}$ which ensures that wave function
ansatz, Eq. (8), be finite for all $r$. Further, for a given $p,$ if $%
a_{p}\neq 0,$ but $a_{p+1}=a_{p+2}=a_{p+3}=\cdots =0,$ we then obtain $%
A_{p}=0$ from Eq. (11), i.e.,
\begin{equation}
E_{p}^{\delta }=c+\sqrt{\frac{2\hbar ^{2}a}{\mu }}\left( 2p+\delta +2\right)
.
\end{equation}
Carrying through a parallel analysis to Ref. [6], $A_{n},$ $B_{n}$ and $%
C_{n} $ must satisfy the determinant relation for a nontrivial solution
\begin{equation}
\det \left|
\begin{array}{cccccc}
B_{0} & C_{1} & \cdots & \cdots & \cdots & 0 \\
A_{0} & B_{1} & C_{2} & \cdots & \cdots & 0 \\
\vdots & \vdots & \vdots & \ddots & \vdots & \vdots \\
0 & 0 & 0 & 0 & A_{p-1} & B_{p}
\end{array}
\right| =0.
\end{equation}
To utilize this method, we present the exact solution for $p=0,1$ as
follows. (1): When $p=0,$ we can obtain from Eq. (15), the exact energy
spectrum as
\begin{equation}
E_{0}^{\delta }=c+\sqrt{\frac{2\hbar ^{2}a}{\mu }}\left( \delta +2\right) .
\end{equation}
Further, it is shown from Eq. (16) that $B_{0}=0$, which leads to the
following restriction on the parameter $\delta $ of the ansatz of wave
function in Eq. (8) with $\eta $ and potential parameters as
\begin{equation}
\delta =-1+\sqrt{\frac{2\mu b}{\hbar ^{2}}+\eta ^{2}}.
\end{equation}
It follows that the wave function for $p=0$ can be written as
\begin{equation}
\psi ^{(0)}(r)=a_{0}r^{\delta +2-D/2}\exp \left[ -\frac{1}{2}\sqrt{\frac{%
2\mu a}{\hbar ^{2}}}r^{2}\right] ,
\end{equation}
where the normalization constant
\begin{equation}
a_{0}=\sqrt{\frac{2}{\left( \delta +1\right) !}}\left( \sqrt{\frac{2\mu a}{%
\hbar ^{2}}}\right) ^{\delta /2+1},
\end{equation}
is determined from the requirement that
\begin{equation}
\int\limits_{0}^{\infty }\left| \psi ^{(n)}(r)\right| ^{2}r^{D-1}dr=1.
\end{equation}
(2): When $p=1,$ the exact energy spectrum becomes
\begin{equation}
E_{1}^{\delta }=c+\sqrt{\frac{2\hbar ^{2}a}{\mu }}\left( \delta +4\right) .
\end{equation}
Furthermore, it is shown from Eq. (16) that $B_{0}B_{1}=A_{0}C_{1},$ which
leads to the restriction on $\delta $ with the parameters of the potential
and $\eta $ as

\begin{equation}
\left( \left( \delta +1\right) ^{2}-\eta ^{2}-\frac{2\mu b}{\hbar ^{2}}%
\right) \left( \left( \delta +3\right) ^{2}-\eta ^{2}-\frac{2\mu b}{\hbar
^{2}}\right) =0.
\end{equation}
The wave function for $p=1$ can be read as

\begin{equation}
\psi ^{(1)}(r)=\left( a_{0}+a_{1}r^{2}\right) r^{\delta +2-D/2}\exp \left[ -%
\frac{1}{2}\sqrt{\frac{2\mu a}{\hbar ^{2}}}r^{2}\right] ,
\end{equation}
where $a_{0}$ and $a_{1}$ are normalization constants. The relation between
them can be determined by Eqs (10)-(14) as

\begin{equation}
4\sqrt{\frac{2\mu a}{\hbar ^{2}}}a_{0}+\left( \left( \delta +3\right)
^{2}-\eta ^{2}-\frac{2\mu b}{\hbar ^{2}}\right) a_{1}=0,
\end{equation}
which provides

\begin{equation}
a_{1}=-\frac{1}{\left( \delta +2\right) }\sqrt{\frac{2}{\left( \delta
+1\right) !}}\left( \sqrt{\frac{2\mu a}{\hbar ^{2}}}\right) ^{\delta /2+2}.
\end{equation}
Following this way, we can generate a class of exact solutions through
setting $p=0,1,2,\cdots ,$ etc. Generally speaking, if $a_{p}\neq 0,$ $%
a_{p+1}=a_{p+2}=\cdots =0,$ we have $A_{p}=0,$ from which we can obtain the
energy spectra \ (cf. Eq. (15)). The wave function can be read

\begin{equation}
\psi ^{(p)}(r)=\left( a_{0}+a_{1}r^{2}+\cdots +a_{p}r^{2p}\right) r^{\delta
+2-D/2}\exp \left[ -\frac{1}{2}\sqrt{\frac{2\mu a}{\hbar ^{2}}}r^{2}\right] ,
\end{equation}
where $a_{i}$ $(i=0,1,2,\cdots ,p)$ are normalization constants.

(i) For the pseudoharmonic potential [8], with the parameters following
Eq.(7), the exact energy spectra become

\begin{equation}
E_{n\ell }=-2D_{e}+\sqrt{\frac{\hbar ^{2}D_{e}}{2\mu r_{e}^{2}}}\left[ 4n+2+%
\sqrt{\frac{8\mu D_{e}r_{e}^{2}}{\hbar ^{2}}+\left[ 2\ell +(D-2)\right] ^{2}}%
\right] ,n,\ell =0,1,2,\cdots
\end{equation}
and radial wave function becomes

\[
\psi ^{(n)}(r)=\left( a_{0}+a_{1}r^{2}+\cdots +a_{n}r^{2n}\right) r^{-\frac{%
(D-2)}{2}+\sqrt{\frac{2\mu D_{e}r_{e}^{2}}{\hbar ^{2}}+\left( \ell +\frac{%
(D-2)}{2}\right) ^{2}}}
\]
\begin{equation}
\times \exp \left[ -\frac{1}{2}\sqrt{\frac{2\mu D_{e}}{\hbar ^{2}r_{e}^{2}}}%
r^{2}\right] .
\end{equation}
(ii) For the three dimensional $(D=3)$ isotropic harmonic oscillator plus
inverse quadratic potential [9], $\ a=\frac{1}{2}\mu \omega ^{2},$ $b=g$ and
$c=0,$ the exact energy spectra are

\begin{equation}
E_{n\ell }=\frac{\hbar \omega }{2}\left[ 4n+2+\sqrt{\left( 2\ell +1\right)
^{2}+\frac{8\mu g}{\hbar ^{2}}}\right] ,n,\ell =0,1,2,\cdots ,g>0
\end{equation}
and the corresponding radial wave function in this case reads

\begin{equation}
\psi ^{(n)}(r)=\left( a_{0}+a_{1}r^{2}+\cdots +a_{n}r^{2n}\right) r^{-1+%
\frac{1}{2}\left[ 1+\sqrt{\left( 2\ell +1\right) ^{2}+\frac{8\mu g}{\hbar
^{2}}}\right] }\exp \left[ -\frac{\mu \omega }{2\hbar }r^{2}\right] .
\end{equation}
where $a_{i}$ with $i=0,1,2,\cdots ,n$ are normalization constants and $%
\omega $ denotes the oscillator frequency. Its worthwhile to note that these
results correspond to the results obtained by Ref. [9].

(iii) Also for $a=B^{2}$ and $b=c=0,$ the potential in Eq. (7) turns to the
anharmonic oscillator potential. And then its exact energy spectra are given
as [8]

\begin{equation}
E_{n\ell }=\sqrt{\frac{\hbar ^{2}}{2\mu }}B\left( 4n+2\ell +D\right) ,n,\ell
=0,1,2,\cdots
\end{equation}
and radial wave function becomes

\begin{equation}
\psi ^{(n)}(r)=\left( a_{0}+a_{1}r^{2}+\cdots +a_{n}r^{2n}\right) r^{\ell
}\exp \left[ -\frac{1}{2}\sqrt{\frac{2\mu }{\hbar ^{2}}}Br^{2}\right] ,
\end{equation}
where $a_{i}$ with $i=0,1,2,\cdots ,n$ are normalization constants. On the
other hand, setting $\hbar =\mu =1,$ $a=\frac{1}{2}$ and $b=c=0,$ gives the
results of Ref. [22] for the results of exact harmonic oscillator energy
states (cf. Eq. (8)) and wave functions (cf. after Eq.(4)).

\subsection{The Mie-type potentials}

This potential has been studied in the $D$ dimensions using the polynomial
solution method [14]. An example on this type of potentials is the standard
Morse [10] or Kratzer-Fues [11,12] potential of the form [10-12]

\begin{equation}
V(r)=-D_{e}\left( \frac{2r_{e}}{r}-\frac{r_{e}^{2}}{r^{2}}\right) ,
\end{equation}
where $D_{e}$ is the dissociation energy between two atoms in a solid and $%
r_{e}$ is the equilibrium internuclear seperation. The standard Kratzer
potential is modified by adding a $D_{e}$ term to the potential. A new type
of this potential is the modified Kratzer-type of molecular potential [15]

\begin{equation}
V(r)=D_{e}\left( \frac{r-r_{e}}{r}\right) ^{2},
\end{equation}
and hence it is shifted in amount of $D_{e}.$ The potential in Eq. (35) will
be studied in $D$ dimensions by making the wave function ansatz [6].
However, this potential has also been discussed before in three dimensions
[15] and in $D$ dimensions [14]. This potential [13] can be simply taken as
\begin{equation}
V(r)=\frac{a}{r}+\frac{b}{r^{2}}+c,
\end{equation}
where $a=-D_{e}r_{e},$ $b=D_{e}r_{e}^{2}$ and $c=D_{e}$ [15]$.$

We take the following ansatz for the radial wave function

\begin{equation}
R(r)=\exp \left[ p\left( \alpha ,r\right) \right] \sum_{n=0}a_{n}r^{n+\delta
+1/2},
\end{equation}
where

\begin{equation}
p\left( \alpha ,r\right) =\alpha r.
\end{equation}
Substituting Eq. (36) into Eq. (4) and setting the coefficient of $%
r^{n+\delta -1/2}$ to zero, we have

\begin{equation}
A_{n}a_{n}+B_{n+1}a_{n+1}+C_{n+2}a_{n+2}=0,
\end{equation}
where

\begin{equation}
A_{n}=-\frac{2\mu a}{\hbar ^{2}}+2\alpha \left( n+\delta +\frac{1}{2}\right)
,
\end{equation}

\begin{equation}
B_{n}=-\frac{2\mu b}{\hbar ^{2}}-\left( \eta ^{2}-\frac{1}{4}\right) +\left(
n+\delta +\frac{1}{2}\right) \left( n+\delta -\frac{1}{2}\right) ,
\end{equation}
\begin{equation}
C_{n}=0,
\end{equation}
and the value of the parameter for $p\left( \alpha ,r\right) $ can be
evaluated as

\begin{equation}
\alpha ^{2}=-\frac{2\mu }{\hbar ^{2}}(E-c),
\end{equation}
from which we must set

\begin{equation}
\alpha =-\sqrt{-\frac{2\mu }{\hbar ^{2}}(E-c)},E<c.
\end{equation}
which ensures that wave function ansatz Eq. (37) be finite for all $r.$ This
is required by the physically acceptable solution. On the other hand, for a
given $p,$ if $a_{p}\neq 0,$ but $a_{p+1}=a_{p+2}=a_{p+3}=\cdots =0,$ it is
easy to obtain $A_{p}=0$ from Eq. (40), i.e.,

\begin{equation}
E_{p}^{\delta }=c-\frac{\mu a^{2}}{2\hbar ^{2}\left( p+\delta +\frac{1}{2}%
\right) ^{2}}.
\end{equation}
To apply this method, we will give the exact solutions for $p=0,1$ below.

(1): When $p=0,$ it is found from Eq. (45) that the exact energy spectrum
becomes

\begin{equation}
E_{0}^{\delta }=c-\frac{\mu a^{2}}{2\hbar ^{2}\left( \delta +\frac{1}{2}%
\right) ^{2}}.
\end{equation}
The restriction on the parameter $\delta $ with potential parameters and $%
\eta $ can be obtained from $B_{0}=0$ as

\begin{equation}
\delta =\sqrt{\frac{2\mu b}{\hbar ^{2}}+\eta ^{2}}.
\end{equation}
The wave function for $p=0$ now becomes

\begin{equation}
\psi ^{(0)}(r)=a_{0}r^{\delta -\frac{(D-2)}{2}}\exp \left[ -\sqrt{-\frac{%
2\mu }{\hbar ^{2}}(E-c)}r\right] ,E<c,
\end{equation}
where

\begin{equation}
a_{0}=\frac{1}{\sqrt{\left( 2\delta +1\right) !}}\left( 2\sqrt{-\frac{2\mu }{%
\hbar ^{2}}(E-c)}\right) ^{\delta +1},
\end{equation}
is a normalization constant obtained via Eq. (21).

(2): When $p=1,$ the exact energy spectrum becomes

\begin{equation}
E_{1}^{\delta }=c-\frac{\mu a^{2}}{2\hbar ^{2}\left( \delta +\frac{3}{2}%
\right) ^{2}}.
\end{equation}
It is shown that the restriction on the parameter $\delta $ with of the
potential parameters and $\eta $ can be obtained from $B_{0}B_{1}=A_{0}C_{1}$

\begin{equation}
\left( \left( \delta +1\right) ^{2}-\eta ^{2}-\frac{2\mu b}{\hbar ^{2}}%
\right) \left( \delta ^{2}-\eta ^{2}-\frac{2\mu b}{\hbar ^{2}}\right) =0.
\end{equation}
The wave function for $p=1$ becomes

\begin{equation}
\psi ^{(1)}(r)=\left( a_{0}+a_{1}r\right) r^{\delta -(D-2)/2}\exp \left[ -%
\sqrt{-\frac{2\mu }{\hbar ^{2}}(E-c)}r\right] ,E<c,
\end{equation}
where the relation between them can be found as

\begin{equation}
-\frac{4\mu a}{\hbar ^{2}}\left( \frac{\delta +1}{\delta +\frac{3}{2}}%
\right) a_{0}+\left( 2\delta +1\right) a_{1}=0,
\end{equation}
which provides

\begin{equation}
a_{1}=\frac{8\mu a}{\hbar ^{2}}\frac{\left( \delta +1\right) }{\left(
2\delta +3\right) \left( 2\delta +1\right) \sqrt{\left( 2\delta +1\right) !}}%
\left( 2\sqrt{-\frac{2\mu }{\hbar ^{2}}(E-c)}\right) ^{\delta +1},E<c.
\end{equation}
Following this way, we can generate a class of exact solutions through
setting $p=0,1,2,\cdots ,$ etc. Generally speaking, if $a_{p}\neq 0,$ $%
a_{p+1}=a_{p+2}=\cdots =0,$ we have $A_{p}=0,$ from which we can obtain the
energy spectra \ (cf. Eq. (45)). The corresponding wave function can also be
read as

\[
\psi ^{(p)}(r)=\left( a_{0}+a_{1}r+\cdots +a_{p}r^{p}\right) r^{-\frac{(D-2)%
}{2}+\sqrt{\frac{2\mu b}{\hbar ^{2}}+\left( \ell +\frac{(D-2)}{2}\right) ^{2}%
}}
\]
\begin{equation}
\times \exp \left[ -\sqrt{-\frac{2\mu }{\hbar ^{2}}(E-c)}r\right] ,E<c
\end{equation}
where $a_{i}$ $(i=0,1,2,\cdots ,p)$ are normalization constants.

(i) For the standard Kratzer-Fues (Mie-type) potential [14], with the
parameters $a=-D_{e}r_{e},$ $b=D_{e}r_{e}^{2}$ and $c=0$, the exact energy
spectra are given as (cf. Ref. [14])

\begin{equation}
E_{n\ell }=-\frac{\hbar ^{2}}{2\mu }\left[ \left( \frac{4\mu D_{e}r_{e}}{%
\hbar ^{2}}\right) ^{2}\left( 2n+1+\sqrt{\frac{8\mu D_{e}r_{e}^{2}}{\hbar
^{2}}+\left[ 2\ell +D-2\right] ^{2}}\right) ^{-2}\right] ,
\end{equation}
and the corresponding radial wave function becomes

\[
\psi ^{(n)}(r)=\left( a_{0}+a_{1}r+\cdots +a_{n}r^{n}\right) r^{-\frac{(D-2)%
}{2}+\frac{1}{2}\sqrt{\frac{8\mu D_{e}r_{e}^{2}}{\hbar ^{2}}+\left( 2\ell
+D-2\right) ^{2}}}
\]
\begin{equation}
\times \exp \left[ -\sqrt{-\frac{2\mu }{\hbar ^{2}}E_{n\ell }}r\right]
,E_{n\ell }<0.
\end{equation}
(ii) For the modified Kratzer potential [15], with the parameters following
Eq. (36), the exact energy spectra are given as (cf. Ref. [15])

\begin{equation}
E_{n\ell }=D_{e}-\frac{\hbar ^{2}}{2\mu }\left[ \left( \frac{4\mu D_{e}r_{e}%
}{\hbar ^{2}}\right) ^{2}\left( 2n+1+\sqrt{\frac{8\mu D_{e}r_{e}^{2}}{\hbar
^{2}}+\left[ 2\ell +D-2\right] ^{2}}\right) ^{-2}\right] ,
\end{equation}
and the radial wave function becomes

\[
\psi ^{(n)}(r)=\left( a_{0}+a_{1}r+\cdots +a_{n}r^{n}\right) r^{-\frac{(D-2)%
}{2}+\frac{1}{2}\sqrt{\frac{8\mu D_{e}r_{e}^{2}}{\hbar ^{2}}+\left( 2\ell
+D-2\right) ^{2}}}
\]
\begin{equation}
\times \exp \left[ -\sqrt{-\frac{2\mu }{\hbar ^{2}}(E_{n\ell }-D_{e})}r%
\right] ,E_{n\ell }<D_{e}.
\end{equation}
(iii) Also after setting $a=-A$ and $b=c=0,$ the potential in Eq. (36) turns
to the Coulomb potential. Then its exact energy spectra are given as [23]

\begin{equation}
E_{n\ell }=-\frac{\mu A^{2}}{2\hbar ^{2}\left( n+\ell +\frac{(D-1)}{2}%
\right) ^{2}},
\end{equation}
and thus the radial wave function reads

\begin{equation}
\psi ^{(n)}(r)=\left( a_{0}+a_{1}r^{2}+\cdots +a_{n}r^{n}\right) r^{\ell
}\exp \left[ -\sqrt{-\frac{2\mu }{\hbar ^{2}}E_{n\ell }}r\right] ,
\end{equation}
where $a_{i}$ with $i=0,1,2,\cdots ,n$ are normalization constants.

\section{Concluding Remarks}

\label{CR}We have easily obtained the exact bound state solutions of the $D$
dimensional radial Schr\"{o}dinger equation for a diatomic molecule with two
general potential forms representing the pseudoharmonic [8] and modified
Kratzer molecular [15] potentials by using the wave function ansatz method
[6]. The presented procedure in this study is systematical and efficient in
finding the exact energy spectra and corresponding wave functions of the
Schr\"{o}dinger equation for various diatomic molecules. This new method is
tested to calculate the energy spectra of the $N_{2},$ $CO,$ $NO$ and $CH$
molecules and our numerical calculations are similar to those given in Table
II with parameter values shown in Table I (cf. Ref. [8]) for pseudoharmonic
potential and also to the ones given by Ref. [15] with the modified Kratzer
potential for the given quantum numbers $n=0,1,2,3,4,5.$ This method is
simple and promising in producing the exact bound state solution for further
anharmonic potentials [6], quarkonium potentials [21] and inter-nuclear
potentials [24].

\section{Acknowledgments}

This research was partially supported by the Scientific and Technological
Research Council of Turkey. S.M. Ikhdair is grateful to Dr. Suat G\"{u}nsel,
founder president of NEU, for a partial fund.

\end{document}